\documentclass[a4paper]{jpconf}
\usepackage{graphicx}
\usepackage{epsfig}
\usepackage{amsfonts}
\usepackage{amssymb}
\usepackage{amsmath}
\usepackage{amsthm}
\usepackage{latexsym}
\usepackage{citesort} 
\usepackage{verbatim} 

\begin{document}
\title{Implications of CP violating 2HDM in B physics}

\author{Anjan S Joshipura, Bhavik P Kodrani}

\address{Physical Research Laboratory, Navarangpura, Ahmedabad 380009, India}

\ead{anjan@prl.res.in, bhavik@prl.res.in}

\def\lsim{\:\raisebox{-1.1ex}{$\stackrel{\textstyle<}{\sim}$}\:}
\def\gsim{\:\raisebox{-1.1ex}{$\stackrel{\textstyle>}{\sim}$}\:}
\def\21{SU(2) $\otimes$ U(1) }
\def\lsim{\raise0.3ex\hbox{$\;<$\kern-0.75em\raise-1.1ex\hbox{$\sim\;$}}}
\def\gsim{\raise0.3ex\hbox{$\;>$\kern-0.75em\raise-1.1ex\hbox{$\sim\;$}}}
\def\vev#1{\left\langle #1\right\rangle}
\newcommand{\ba}{\begin{array}}
\newcommand{\ea}{\end{array}}
\newcommand{\be}{\begin{equation}}
\newcommand{\ee}{\end{equation}}
\newcommand{\beqa}{\begin{eqnarray}}
\newcommand{\eeqa}{\end{eqnarray}}
\newcommand{\bsbsbar}{$B^0_s-\bar{B}^0_s~~$}
\newcommand{\bqbqbar}{$B^0_q-\bar{B}^0_q~~$}
\newcommand{\kkbar}{$K^0-\bar{K}^0~$}
\newcommand{\bdbdbar}{$B^0_d-\bar{B}^0_d~~$}

\begin{abstract}
The charged fermion mass matrices are invariant under $U(1)^3$
symmetry linked to the fermion number transformation. Under the condition that
the definition of this symmetry in arbitrary weak basis does not depend upon
Higgs parameters such as ratio of vacuum expectation values, a class of two
Higgs doublet models (2HDM) can be identified in which tree level flavor
changing neutral currents normally present in 2HDM are absent. However unlike
the type I or type II Higgs doublet models, the charged Higgs couplings in
these models contain additional flavor dependent CP violating phases. These
phases can account for the recent hints of the beyond standard model CP
violation in the $B_d$ and $B_s$ mixing. In particular, there is a range of
parameters in which new phases do not contribute to the $K$ meson CP
violation but give identical new physics phases in the $B_d$ and
$B_s$ meson mixing.
\end{abstract}

\section{Introduction}
Cabibbo Kobayashi Maskawa (CKM) mechanism of CP violation has been established
as the dominant source of CP violation by observations
at B-factories and Tevatron. These experiments have also shown some hints of
physics beyond standard model
\cite{Bona:2009tn,Botella:2006va,Lenz:2007nk,Lenz:2006hd,Nierste:2006na}. One
of them is CP violating observable $S_{J/\psi
K_s}$ of \bdbdbar mixing which is measured in time dependent CP
asymmetry of $B_d^0 \rightarrow J/\psi K_s$ by BABAR experiment at
the Stanford linear accelerator center (SLAC) and by Belle
experiment at KEK \cite{:2009yr,Chen:2006nk}. Value of
$S_{J/\psi K_s}$ obtained from fit using $\epsilon_k, \Delta m_q,
|V_{cb}|,\alpha, \gamma$ as inputs
shows deviations from experimentally determined value at about $2\sigma$
\cite{Lunghi:2009sm}.
Another hint for new physics comes from determination of CP
violating phase $\phi_s$ of \bsbsbar mixing. $\phi_s$ is determined
from analysis of time dependent angular distribution of decay
products in flavor tagged decay $B_s^0 \rightarrow J/\psi \phi$
 by CDF and D0 experiments at Fermilab Tevatron collider
through decay chain $B_s^0 \rightarrow J/\psi \phi$, $J/\psi
\rightarrow \mu^+ \mu^-$, $\phi \rightarrow K^+
K^-$ \cite{Aaltonen:2007he,:2008fj}. SM predicted
value for  $\phi_s$ is much
small compared to the experimentally determined value. UTfit group has combined
all the available experimental constraints on $B_s$ mixing and performed a model
independent analysis of NP contribution to \bsbsbar mixing. From this analysis
UTfit group has concluded that phase $\phi_s$ of \bsbsbar mixing
determined from their analysis deviates from the SM prediction by
about 3$\sigma$ \cite{Bona:2008jn,Bona:2009tn}. A
similar analysis performed by CKMfitter group shows that the deviation from SM
prediction is about 2.5 $\sigma$ \cite{Deschamps:2008de}.

Inconsistencies between SM prediction and the experimental
determination in \bdbdbar and \bsbsbar mixing discussed here can be
resolved if there is some new physics beyond SM which can give extra
contribution to above mentioned CP violating observables.  Here we
explore the possibility of explaining these deviations in two Higgs
doublet model (2HDM). The most general 2HDM generates large flavor
changing neutral currents (FCNC) for moderate Higgs mass.
To avoid FCNC an additional discrete symmetry is imposed which results in
type-I or type II 2HDM. In these models the charged Higgs couplings
do not provide any additional source of CP violation. Therefore they
cannot explain possibly large CP violating phases in the neutral
$B$ mixings if confirmed in future. Hence we need to go beyond the
type-I and type-II 2HDM to explain new CP violating phases. One way
is to restrict the structure of FCNC couplings using flavor
symmetries rather than eliminating them altogether 
\cite{Antaramian:1992ya,Cheng:1987rs,Branco:1996bq,Hall:1993ca,Joshipura:1990xt}
. This leads to models with suppressed FCNC and additional phases whose
phenomenological consequences have been studied in
\cite{Joshipura:2009ej,Joshipura:2007sf,Joshipura:2007cs}. A different
aproach based on shared flavour symmetry is given in \cite{Datta:2008qn}.
An intersting possibility is to have models without tree level
FCNC but containing additional phases in the charged Higgs couplings. Here we
wish to discuss such models motivated by the studies of flavor
symmetries of mass matrix in 2HDM.

Flavor symmetries are often applied to restrict the structure of
Yukawa couplings all of which cannot be directly determined from
experiments.  The relation between the structures of mass
matrices and symmetries was studied in 
\cite{Grimus:2009pg,Lam:2008rs,Lam:2008sh,Lam:2006wm}. It
was shown by Lam that a symmetry can always be found under which an arbitrary
neutrino mass matrix $M_\nu$ remains invariant \cite{Lam:2006wm}. Grimus,
Lavoura and Ludl \cite{Grimus:2009pg} showed that any Hermitian mass matrix $M_f
M_f^\dagger$ obtained form a fermion mass matrix $M_f$ always
possesses a symmetry $G_f=U(1)\times U(1)\times U(1)$ and the
corresponding $G$ for the mass matrix of the Majorana neutrinos is
$Z_2\times Z_2\times Z_2$. Here we generalize this study to the non
hermitian mass matrices in 2HDM.

In the next section we will describe general 2HDM and the structure of
FCNC and charged higgs interaction. Study of the symmetries of mass matrices in
2HDM and their consequences on FCNC and charged higgs interaction will also be
given. In sction 3. phenomenology of this model in neutral mesons  mixing
will be given. In section 4. we will obtain constraints on NP paramateres in
present case from the neumerical analysis of \bsbsbar mixing and charge
asymmetry of like sign dimuon events. Last section will present a summary.

\section{Mass matrix symmetries and 2HDM}
 Yukawa couplings for 2HDM are given as
\be \label{yukawa1} -{\cal
L}_Y=\bar{Q}_{L}'(\Gamma_{1d}\phi_1+\Gamma_{2d}\phi_2)d_R'+\bar{Q}_{L}'(\Gamma_{1u}\tilde{\phi}_1+
\Gamma_{2u}\tilde{\phi}_2)u_R'+{\rm H.c.} ~,\ee where
$\Gamma_{iq}~~(i=1,2;q=u,d)$ are matrices in the generation space.
$\phi_{1,2}$ denote  Higgs doublets and $\tilde{\phi}_i= i
\tau_2\phi_i^*$. $Q'_{iL}$ refer to three generations of doublet
quarks and primed fields in the above  equation refer to various
quark fields in the weak basis. The neutral component of a specific
linear combination of the Higgs fields $\phi\equiv \cos\beta \phi_1+\sin\beta  e^{-i
\theta}\phi_2$ is responsible for the mass generation
 \be
\label{mq} M_q=v (\cos\beta \Gamma_{1q}+\sin\beta \Gamma_{2q}
e^{i\theta_q})=V_{qL} D_qV_{qR}^\dagger~,\ee where
$\vev{\phi_1^0}=v\cos\beta~;~\vev{\phi_2^0}=v\sin\beta e^{i\theta}$ with $v\sim 174$
GeV and $\theta_d=-\theta_u=\theta$. The matrices
$V_{qL,R}$ diagonalize $M_q$. We define the matrices $S_{qL,R}$ as
\be \label{sqlr}S_{qL,R}=V_{q L,R}P_q V_{q L,R}^\dagger ~,\ee
It can be seen that the mass matrices have invariance given by following
equation. \be \label{Sq} S_{qL}^\dagger M_q S_{qR}= M_q~.\ee Where $P_q={\rm
diag.}(e^{i\alpha_{1q}},e^{i\alpha_{2q}},e^{i\alpha_{3q}})$. Here
$S_{qL,R}$ define two different $U(1)\times U(1)\times U(1)$
symmetries $G_u$ and $G_d$ for up and down quarks. We put a mild requirement on
possible $S_{q L,R}$ namely that the form of $S_{q L,R}$ be
independent of the parameters $\tan\beta$ and $\theta$ which are
determined  entirely in the Higgs sector.  With this requirement and
from eq.(\ref{mq}) of mass matrices, we get
 \be
\label{gamaq}
S_{qL}^\dagger \Gamma_{iq}S_{qR}=\Gamma_{iq}~~~~~{i=1,2} ~. \ee This
shows that individual Yukawa
couplings should also  respect the symmetry. Let us parametrize
$\Gamma_{iq}$ as \be \label{paramet} \Gamma_{iq}\equiv V_{qL}
\tilde{\Gamma}_{iq}V_{qR}^\dagger~.\ee Eqs. (\ref{sqlr},\ref{gamaq})
then imply \be \label{sol} P_q^\dagger\tilde{\Gamma}_{iq}
P_q=\tilde{\Gamma}_{iq} ~. \ee If $G_u,G_d$ refer to the full
$U(1)\times U(1)\times U(1)$ symmetry with totally independent
$\alpha_{iq}$ then the  only non-trivial solution of eq.(\ref{sol})
is a diagonal $\tilde{\Gamma}_{iq}$ for every $i$ and $q$. Yukawa
couplings are then given as \be \label{yukawa2} \Gamma_{iq}=V_{qL}
\gamma_{iq}V_{qR}^\dagger  ~,\ee where $\gamma_{iq}$ are diagonal
matrices with  complex entries. More general forms for
$\tilde{\Gamma}_{iq}$ are allowed if one demands invariance with
respect to subgroups of $G_u\times G_d$. All the flavor violations
are induced by Higgs combination $\phi_F\equiv -\sin\beta
\phi_1+\cos\beta \phi_2 e^{-i\theta}$. The couplings of the neutral component
$\phi_F^0$ are given as \be \label{fcnc} {-\cal
L}_Y^0=\bar{q}_{L}F_q q_{R}\phi_F^0+{\rm H.c.} \ee with
\begin{eqnarray} \label{fij}
F_q &\equiv& V_{qL}^\dagger (-\sin\beta \Gamma_{1q}+\cos\beta \Gamma_{2q} e^{i
\theta_q})V_{qR} \nonumber \\  &=& (-\sin\beta \gamma_{1q}+\cos\beta \gamma_{2q}
e^{i\theta_q})~,
\end{eqnarray}
 where we have used eq.(\ref{yukawa2}) to obtain the second line. It is seen
that the FCNC matrix $F_q$ become diagonal along with the mass matrices and the
tree level FCNC are absent. But the phases of $(F_q)_{ii}$ cannot be removed in
the process of making the quark masses real and remain as physical
parameters. The charged component $H^+$ of $\phi_F$ correspond to the
physical charged Higgs field and its couplings in our case are given
by \be \label{hplus} -{\cal L}_{H^+}=\frac{H^+}{v} \left(
\bar{u}_{iL} V_{ij} (F_d)_{jj} d_{jR} - \bar{u}_{iR} V_{ij}
(F_u^*)_{ii} d_{jL}\right)+{\rm H.c.} ~\ee
The above couplings  are similar to the charged Higgs couplings in
2HDM of type-I and II. In those models, $(F_q)_{ii}$ are
proportional to the corresponding quark masses $m_{iq}$ and are
real. Here $(F_q)_{ii}$ are general complex numbers which can
provide new phases in the  $B_{d,s}-\bar{B}_{d,s}$ mixing.

An interesting class of 2HDM without the tree level
FCNC have been obtained from general 2HDM by assuming that two Yukawa couplings
$\Gamma_{1q}$ and $\Gamma_{2q}$ are proportional to each other 
\cite{Pich:2009sp}.  In the present case,  the two Yukawa coupling matrices are
not proportional to each other but the tree level FCNC are still
absent. The phases of $(F_q)_{ii}$ in the charged Higgs couplings
are dependent on the flavour index $i$ unlike in models of
\cite{Pich:2009sp} which are characterized by universal phases one
for the up and the other for the down quarks. If the diagonal matrices
$\gamma_{1q}$ and $\gamma_{2q}$ in eq.(\ref{yukawa2}) are
proportional to each other then the present class of models reduce to the one in
\cite{Pich:2009sp}.

If $S_{uL}\not = S_{dL}$ then neither the Yukawa
interactions given in eq.(\ref{yukawa1}) nor the charged current weak
interactions remain invariant under symmetries of the mass matrices.
This means that radiative corrections will not preserve
\cite{Ferreira:2010xe} the structure implied by eq.(\ref{yukawa2}). However as
in case of the 2HDM of type-I and type-II as well as the aligned models of
\cite{Pich:2009sp}, 
the full Lagrangian of the present model is formally invariant under
the  fermion number transformation $q_{iL,R}\rightarrow
e^{i\alpha_{qi}}q_{iL,R}$ accompanied by the change in the CKM
matrix elements $V_{ij}\rightarrow e^{i \alpha_{iu}}V_{ij}e^{-i
\alpha_{jd}}$. As a consequence of this all the radiative
corrections in the model would display structure similar to the one
obtained in the Minimal Flavor Violating \cite{D'Ambrosio:2002ex} models.\\

\section{Neutral Mesons mixings}
In this section we discuss phenomenological application of the model in
new contribution to neutral meson mixing induced by the charged Higgs
couplings in eq.(\ref{hplus}). The $(F_d)_{ii}$ and $(F_u)_{ii}$ entering the
$H^+$ couplings are determined by the diagonal Yukawa couplings
$\gamma_{iq}$ which also determine corresponding quark masses, see
eq.(\ref{yukawa2}). We make a simplifying assumption that the
first two generation quark masses and the corresponding $(F_q)_{ii}$
are small compared to the third generation masses and  $(F_q)_{33}$.
Hence eq.(\ref{hplus}) reduces to
\be
\label{approxhplus} -{\cal L}_{H^+}\approx\frac{H^+}{v} \left(
\bar{u}_{iL} V_{i3} (F_d)_{33}b_R - \bar{t}_{R} V_{3j}
(F_u^*)_{33}d_{jL}\right)+{\rm H.C.} ~\ee
The charged Higgs contribution to the $K^0-\bar{K}^0$ mixing arise only from the
second term. The phase in $(F_u)_{33}$  can be absorbed in the
definition of $H^+$. As a result the above Lagrangian does not
generate any new CP violating phases in the $K$ meson mixing as long
as $(F_q)_{jj}$ are neglected for $j=1,2$. However it can lead to non-trivial
phases in the $B_q-\bar{B}_q$ mixing since the charged Higgs
exchanges in this case involve both $(F_d)_{33}$ and $(F_u)_{33}$ and their
phases can not be simultaneously removed. More interestingly, the
interaction in eq.(\ref{approxhplus}) distinguishes between the $d$ and $s$
quarks only through the CKM factor and not through additional phases. This
results in strong correlations among the CP violation in $B_s$ and
$B_d$ system. This can be seen as follows.

$B_q^0-\bar{B}_q^0$  ($q=d,s$) mixing amplitude can be parameterized
in the presence of new physics contribution as
 \be \label{np}
 \vev{B_q|{\cal H}^{SM}+{\cal H}^{NP}|\bar{B}_q}\equiv
\vev{B_q|{\cal H}^{SM}|\bar{B}_q}(1+\kappa_q e^{i \phi^{NP}_q})\equiv
|\vev{B_q|{\cal H}^{SM}|\bar{B}_q}|\rho_q e^{-2 i
(\beta_q+\phi_q)}~, \ee where $\beta_q$ represent the relevant phase
in case of the SM and $\phi_q$ are the charged Higgs induced phases.
The box diagrams involving $WH$ and $HH$ lead to new physics
contribution involving the charged Higgs $H$. The interaction in
eq.(\ref{approxhplus}) lead to the following effective Hamiltonian at the weak
scale:
 \be \label{effective H} {\cal H}^{NP}=C_1\bar{q_L}\gamma^\mu b_L
\bar{q_L}\gamma_\mu b_L+C_2\bar{q_R} b_L \bar{q_R}b_L ~,\ee
where $C_{1,2}$ are the Wilson coefficients. Taking matrix element of the
above equation and comparing with the SM result leads to
\be\label{kapaq} \kappa_q e^{i\phi_q^{NP}}=\frac{4 \pi^2}{G_F^2
M_W^2\eta_B S_0(x_t)}\left[C_1-5/24 C_2 \left(
\frac{M_{B_q}}{m_b+m_q}\right)^2\frac{B_{2q}}{B_{1q}}\right]  ~,\ee
where $G_F,M_W$ respectively denote the Fermi coupling constant and
the W boson mass. $S_0(x_t)\approx 2.3 $ for $m_t\sim 161 $ GeV.
$\eta_B\approx 0.55$ refers to the QCD correction to the Wilson
operator in the SM, $B_{1q,2q}$ are the bag factors which enter the
operator matrix elements in eq.(\ref{effective H}) and
$M_{B_q},m_b,m_q$ respectively denote the masses for the $B_q$
mesons for $q=d,s$, b quark and the $d,s$ quarks. The Wilson coefficients
$C_{1,2}$ are independent of the flavour $q=d,s$ of the light quark in $B_q$.
Mild dependence of $\kappa_q$ on $q$ arise from the operator matrix element
multiplying $C_2$ in eq.(\ref{kapaq}). This leads to two predictions: To a good
approximation, ($i$) $\kappa_d \approx \kappa_s$ and ($ii$)
$\phi_{d}^{NP}\approx \phi_s^{NP}$. This implies from eq.(\ref{np})
that \be \label{ratio} \frac{\Delta M_d}{\Delta M_s}\approx
\frac{\Delta M_d^{SM}}{\Delta M_s^{SM}}~,\ee where $\Delta M_q$
denote the values of the mass difference between heavy and
light mass eigenstates in $B_q$ system in presence of new physics. Equality of
$\kappa_d$ and $\kappa_s$ as well as $\phi_d^{NP}$ and
$\phi_s^{NP}$ along with eq.(\ref{np}) also imply
\be \label{phases} \phi_d\approx \phi_s~ \ee The detailed
phenomenological consequences of this prediction are already
discussed in \cite{Buras:2008nn} in a model independent manner. It
appears to
be in the right direction for explaining the CP violating anomalies.
In case of $B_d$, the unitarity triangle angle $\beta$
$$ \sin 2\beta=0.84\pm 0.09$$ as determined \cite{Lunghi:2009sm} using
the information from
$V_{cb},\epsilon_K$  and $\frac{\Delta M_d}{\Delta M_s}$ is found to be
higher than the value
$$ \sin 2\beta=0.681\pm 0.025$$
obtained from the mixing induced asymmetry in $B\rightarrow J/\psi
K_S$ decay. Since the latter measures $\beta+\phi_d$, the above
information can be reconciled \cite{Buras:2008nn}  with a negative
$\phi_d\approx -10^\circ$. $\phi_d\approx \phi_s$ then implies a
sizable asymmetry \cite{Buras:2008nn}  $S_{\psi\phi}=\sin
2(\beta_s-\phi_s)\sim 0.4 $ in $B_s$ decay as indicated by the
analysis of UTfit \cite{Bona:2008jn} or the CKMfitter
\cite{Deschamps:2008de} group. 

In a recent study global fit to various obseravbles was carried out in
different scenarios \cite{Lenz:2010gu}. One of the scenario has the relation
$\kappa_d \equiv \kappa_s$ and $\phi_d^{NP} \equiv \phi_s^{NP}$. The resulting
fit in this scenario is better than the SM fit. Results from this fit are a
comman NP phase 
$2\phi_d = 2 \phi_s = -14.4^{+6.7}_{-4.2}$ and $\sin 2\beta = 0.83 \pm 0.05$
at $2\sigma$. These results agree with our results obtained here. 
\section{Numerical analysis}
\begin{figure}[h]
\begin{center}
  \includegraphics[width=6.0cm]{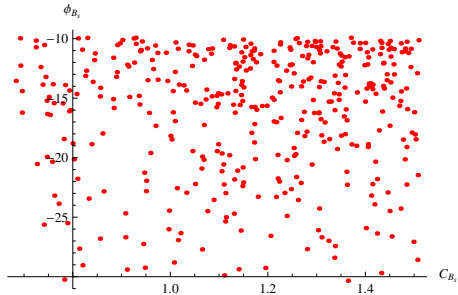}
\end{center}
  \caption{\label{fig:NPUTfit} Values of NP parameter $C_{B_s}$ and $\phi_{B_s}$
which can be obtained in limit when first two generation masses and
 $F_{qii} ~ ,\text{where} ~ q = u,d$ and $i = 1,2 $, can be
neglected}
\end{figure}

In this section we discuss our numerical analysis in which
we obtain constraints on model parameters from study of \bsbsbar mixing. New
physics contribution to \bsbsbar mixing is parametrized by
UTfit group as \cite{Bona:2008jn}
\begin{equation}
C_{B_S} e^{-2 i \phi_{B_s}} = \left( 1 + \frac{\Delta M_s^{NP}}{\Delta
M_s^{SM}}\right)
\end{equation}
Also from eq.(\ref{np}), $C_{B_s} = |1 + \kappa_s e^{-2i (\phi_s^{NP}-\beta_s)} |$ and $\phi_{B_S} = -\frac{1}{2} \text{Arg}[1 + \kappa_s e^{-2i
(\phi_s^{NP}-\beta_s)}]$. UTfit collaboration has obtained the allowed range of
the parameters $C_{B_S}$ and $\phi_{B_s}$ as \cite{Bona:2008jn}
\begin{eqnarray} \label{UTfitvalues}
C_{B_s} &=& [0.68 , 1.51] ~, \nonumber \\
\phi_{B_s} &=& [-30.5 , -9.9] \cup [-77.8 , -58.2]
\end{eqnarray}
We calculate $C_{B_s}$ and $\phi_{B_s}$ in the present
model neglecting first and second generation masses and corresponding $F_{qii}$.
The Charged higgs mass $M_h$ is allowed to vary in the range $100-500$ GeV while
$\beta$ and couplings $F_{q33}$ are varied in the allowed range. Values of
$C_{B_s}$ and $\phi_{B_s}$ which can be obtained in this case is shown in
Figure(\ref{fig:NPUTfit}).

We have also checked that our model is compatible with the recent
determination of like-sign dimuon charge asymmetry of semileptonic
b-hadron decays by D0 collaboration \cite{Abazov:2010hv}. The model
discussed here can give extra contributation required to explain the deviation
of the experimental value of like-sign dimuon charge
asymmetry of semileptonic b-hadron decays from SM prediction. 
\section{Summary}
Using the flavor symmetries which does not depend on the higgs
parameter $\tan \beta$ and $\theta$, we obtained a class of 2HDM in
which FCNCs can be eliminated without imposing discrete symmetries.
Unlike type - I and type-II 2HDM, charged higgs interaction in this
model contains new phases which are flavor dependant. In the limit of
vanishing first and second generation masses and corresponding couplings
$F_{qii} ~ \text{where} ~ q = u,d$ and $i = 1,2 $, \kkbar mixing does not
get any new CP phase while the \bdbdbar and \bsbsbar mixing gets new phases. New
contribution to CP violation in this model in \bdbdbar and \bsbsbar mixing are
correlated and it is possible to explain CP violating anomalies in $B_d$ and
$B_s$ systems. The model parameters can also generate like sign dimuon charge
asymmetry required by the experimental data.
\section{References}
\bibliographystyle{iopart-num} 
\bibliography{bib-PASCOS}
\end{document}